\documentclass{article}
\usepackage{dcase2021,amsmath,graphicx,url,times,booktabs, tabularx}
\usepackage{amssymb,amsfonts}
\usepackage{multirow}
\graphicspath{}

\title{Continual Learning for Automated Audio Captioning Using The Learning Without Forgetting Approach}
\name{Jan Berg and Konstantinos Drossos}
\address{Audio Research Group, Tampere University, Finland\\\{firstname.lastname\}@tuni.fi\vspace{6pt}
 }

\begin{document}

\ninept
\maketitle

\begin{sloppy}

\begin{abstract}
Automated audio captioning (AAC) is the task of automatically creating textual descriptions (i.e. captions) for the contents of a general audio signal. Most AAC methods are using existing datasets to optimize and/or evaluate upon. Given the limited information held by the AAC datasets, it is very likely that AAC methods learn only the information contained in the utilized datasets. In this paper we present a first approach for continuously adapting an AAC method to new information, using a continual learning method. In our scenario, a pre-optimized AAC method is used for some unseen general audio signals and can update its parameters in order to adapt to the new information, given a new reference caption. We evaluate our method using a freely available, pre-optimized AAC method and two freely available AAC datasets. We compare our proposed method with three scenarios, two of training on one of the datasets and evaluating on the other and a third of training on one dataset and fine-tuning on the other. Obtained results show that our method achieves a good balance between distilling new knowledge and not forgetting the previous one. 
\end{abstract}


\section{Introduction}
\label{sec:intro}
Automated audio captioning (AAC) is the inter-modal translation task, where a method takes as an input a general audio signal and generates a textual description of the contents of the audio signal~\cite{drossos2017automated}. AAC methods learn to describe sound sources/events, spatiotemporal relationships of events, textures and sizes, and higher-level knowledge like counting~\cite{drossos2017automated,koizumi:2020:interspeech}, but not speech transcription~\cite{drossos2019clotho,lipping2019dcase}. In a typical AAC scenario, a deep learning method is optimized in a supervised or reinforcement learning scheme and using an AAC dataset~\cite{Takeuchi2020,Cakir2020,Xu2020,Chen2020,Nguyen2020a}. Audio clips are given as an input to the AAC method and the method generates captions for its inputs. Then, the method is optimized by trying to reduce the difference between the predicted and the actual (i.e ground truth) captions. Given that the existing AAC datasets are limited, the above scheme creates some limitations. For example, since the available information from the audio clips in the different datasets are most likely not overlapping and the described information and expression variability differs given that different annotators have been used~\cite{drossos2019clotho,audiocaps}, then an AAC method optimized with one dataset will have problems when evaluated with another AAC dataset. Even if some technique is used for adapting an AAC method to another dataset, e.g. like transfer learning, it would be required to have all the new data for the adaptation. This creates limitation of continuously adapting an AAC method to new information.

The above presented problem of continuously adapting is not new and has been attacked using continual learning, sometimes also called lifelong learning~\cite{PARISI201954, 10.5555/3276462}, which is the process of continuously adapting a method to new data and/or tasks. The advantage of continual learning over other techniques, e.g. transfer learning, is that the latter usually introduces the phenomenon of catastrophic forgetting, where the method is adapted to the new information but forgets the initially learned one~\cite{PARISI201954, FRENCH1999128,kirkpatrick2017overcoming, li2017learning}. Though, continual learning methods seem that tackle this phenomenon~\cite{kirkpatrick2017overcoming, li2017learning}. There are different approaches for continual learning, e.g. like joint training~\cite{10.5555/3276462}, though our focus is on the cases where the new data are not required a priori, because it is often not possible to have all data beforehand due to storing reasons (e.g. cannot store all the data) or to degradation of data (e.g. data have been lost over time). Approaches that do not require having the data to do the adaptation, can be roughly divided into three categories~\cite{PARISI201954}, namely reguralization methods like learning without forgetting (LwF)~\cite{li2017learning} and elastic weight consolidation (ECW)~\cite{kirkpatrick2017overcoming}, dynamic architectures like dynamically expandable networks (DEN)~\cite{yoon2018lifelong}, and replay models like gradient episodic memory (GEM)~\cite{lopez2017gradient}.

In this paper we consider the scenario where an AAC method continuously adapts to new and unseen data, using unseen ground truth captions. This scenario can resemble, for example, an online platform where new audio data and captions can be provided by human users and the AAC method continuously learn from the new data. Focusing on this, we present a first method for continual learning for AAC, adopting the LwF approach. Although there are published continual learning approaches for audio classification using different approaches~\cite{tao2020fewshot,9413584}, we employ LwF due to its simplicity, reduced need for resources, and the facts that LwF is model agnostic and no modifications are needed for the employed AAC model. Although, previous research has shown that one of the weaknesses of LwF is that its effectiveness is dependent on the similarity of the tasks at hand~\cite{aljundi2017expert, PARISI201954, Delange_2021}, we deem that this is not applicable to our case since we use LwF for continuously adapting to new data on the same task. 

For our work presented here, we employ a freely available and pre-optimized AAC method called WaveTransformer (WT)~\cite{tran2020wavetransformer} and two freely available AAC datasets, namely Clotho~\cite{drossos2019clotho} and AudioCaps~\cite{audiocaps}. Since WT method has achieved state-of-the-art results on Clotho, we use AudioCaps as the new data that the AAC method will adapt. Given that there are no other published continual learning approaches for AAC, in this paper we do not consider the case of the mismatched set of words in the two employed datasets. The rest of the paper is organized as follows. Section~\ref{sec:method} presents our method and Section~\ref{sec:evaluation} presents the adopted evaluation process. Obtained results are in Section~\ref{sec:results} and Section~\ref{sec:conlusions} concludes the paper. 

\section{Method}\label{sec:method}
Our method is model agnostic, based on LwF and knowledge distillation~\cite{li2017learning,hinton2015distilling}. It employs a pre-optimized AAC model, a copy of the AAC model, an iterative process, a regularization-based loss, and a stream of new audio data with captions that are used for learning the new information. The stream of new audio data and captions is used to provide input to the original and copy AAC models. The output of both models is used against the provided captions from the stream, but only the parameters of the copy model is updated. At every update of the parameters, the copy model can be used as an output of our continual learning method. An illustration of our continual learning approach is in Figure~\ref{fig:method}.

In more detail, we start by having a pre-optimized AAC model $\text{M}_{\text{base}}(\cdot;\theta_{\text{base}})$, having the pre-optimized parameters $\theta_{\text{base}}$. $\text{M}_{\text{base}}$ is pre-optimized using a dataset of $K$ input-output examples $\mathbb{D}_{\text{ori}} = \{(\mathbf{X}', \mathbf{Y}')_{k}\}_{k=1}^{K}$, where $\mathbf{X}'\in\mathbb{R}^{T_{\text{a}}\times F}$ is a sequence of $T_{\text{a}}$ audio feature vectors having $F$ features and $\mathbf{Y}'\in\{0, 1\}^{T_{\text{w}}\times W}$ a sequence of $T_{\text{w}}$ one-hot encoded vectors of $W$ elements, that indicate the most probable word for each $t_{\text{w}}$-th word index. $\text{M}_{\text{base}}$ generates its output as 

\begin{equation}
    \hat{\mathbf{Y}}'_{k} = \text{M}_{\text{base}}(\mathbf{X}'_{k};\theta_{\text{base}})\text{,}
\end{equation}
\noindent
where $\hat{\mathbf{Y}}'_{k}$ is the predicted caption by $\text{M}_{\text{base}}$ when having as input the $\mathbf{X}'_{k}$. The optimization of $\theta_{\text{base}}$ is performed by minimizing the loss

\begin{equation}
    \mathcal{L}(\theta_{\text{base}}, \mathbb{D}_{\text{ori}}) = \sum\limits_{k=1}^{K}\text{CE}(\mathbf{Y}'_{k}, \hat{\mathbf{Y}}'_{k})\text{,}
\end{equation}
\noindent
where $\text{CE}$ is the cross-entropy loss between $\mathbf{Y}'_{k}$ and $\hat{\mathbf{Y}}'_{k}$. 

Then, we create a copy of $\text{M}_{\text{base}}$, $\text{M}_{\text{new}}(\cdot;\theta_{\text{new}})$, having same hyper-parameters as $\text{M}_{\text{base}}$ and the parameters $\theta_{\text{new}}$. Our target is to continuously update $\theta_{\text{new}}$ for new data, without making $\text{M}_{\text{new}}$ to deteriorate its performance on $\mathbb{D}_{\text{ori}}$. The new data are coming from a stream of data, $\mathcal{S}$, which continually produces new and unseen data (i.e. data not in $\mathbb{D}_{\text{ori}}$). We sample data from $\mathcal{S}$ in batches of $B$ examples, creating the input-output examples as
\begin{equation}\label{eq:data-sampling}
    \mathbb{D}_{\text{new}} =\{(\mathbf{X}, \mathbf{Y})_{b}: (\mathbf{X}, \mathbf{Y})\sim\mathcal{S}\wedge b=1,\ldots, B\}\text{,}
\end{equation}
\noindent
where $\mathbf{X}\in\mathbb{R}^{T_{\text{a}}\times F}$ is a sequence of audio features, similar to $\mathbf{X}'$, and $\mathbf{Y}\in\{0, 1\}^{T_{\text{w}}\times W}$ is a sequence of one-hot encoded vectors similar to $\mathbf{Y}'$. Here has to be noted that the captions coming from $\mathcal{S}$ can (and most likely will) have different set of words with $\mathbf{Y}'$. Though, our approach is not considering the problem of the different set of words. For that reason, we consider from $\mathbf{Y}$ only the words that are common with $\mathbf{Y}'$. 

\begin{figure}
    \centering
    \includegraphics[width=\columnwidth]{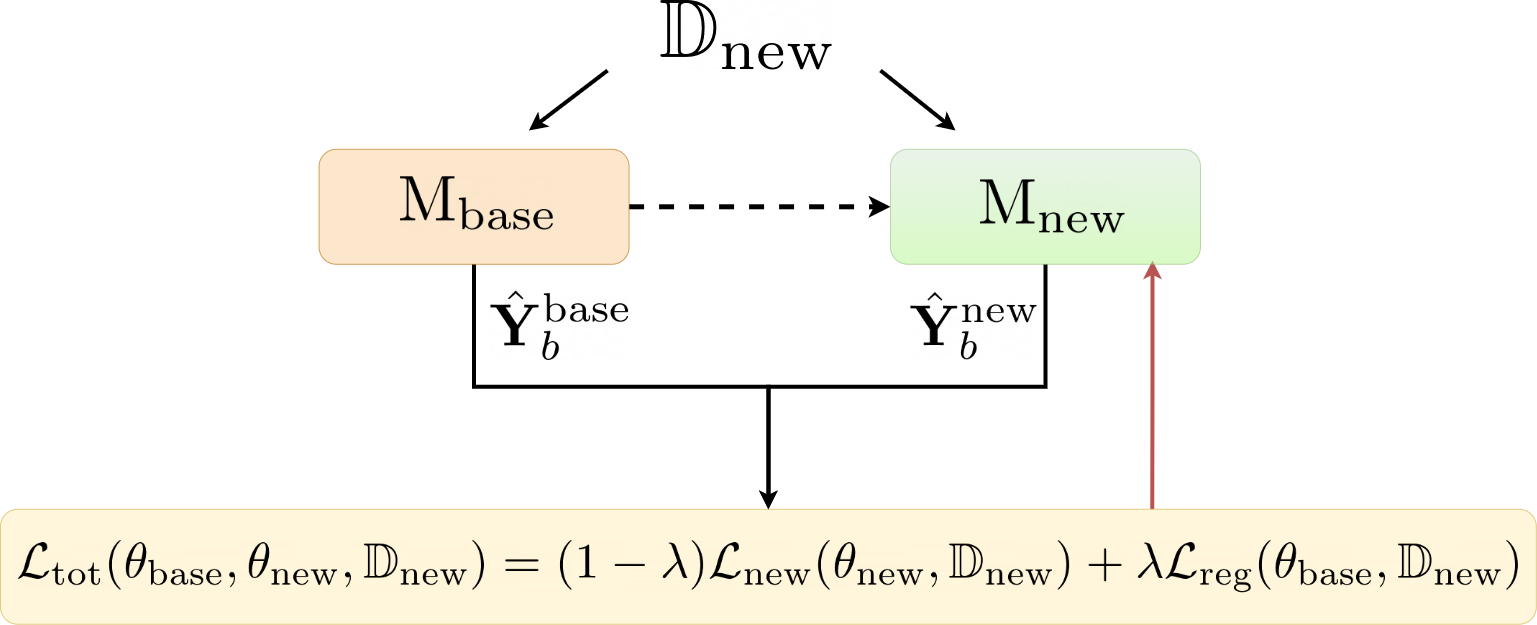}
    \caption{Our proposed continual learning method for AAC. The dotted line represents the copying of the parameters of $\text{M}_{\text{base}}$ to $\text{M}_{\text{new}}$, and it takes place only once at the beginning of the process. Red line indicates backpropagation for updating the parameters of $\text{M}_{\text{new}}$.}
    \label{fig:method}
\end{figure}

We use the sampled data $\mathbb{D}_{\text{new}}$ as an input to both $\text{M}_{\text{base}}$ and $\text{M}_{\text{new}}$, resulting to

\begin{align}
    \hat{\mathbf{Y}}^{\text{base}}_{b} &= \text{M}_{\text{base}}(\mathbf{X}_{b};\theta_{\text{base}})\text{, and}\\
    \hat{\mathbf{Y}}^{\text{new}}_{b} &= \text{M}_{\text{new}}(\mathbf{X}_{b};\theta_{\text{new}})\text{,}
\end{align}
\noindent
where $\hat{\mathbf{Y}}^{\text{base}}_{b}$ and $\hat{\mathbf{Y}}^{\text{new}}_{b}$ are the predicted outputs of $\text{M}_{\text{base}}$ and $\text{M}_{\text{new}}$, respectively, when having as an input $\mathbf{X}_{b}$. 

Having $\hat{\mathbf{Y}}^{\text{base}}_{b}$ and $\hat{\mathbf{Y}}^{\text{new}}_{b}$, we define the loss

\begin{align}
    \label{eq:total-loss}
    \mathcal{L}_{\text{tot}}(\theta_{\text{base}}, \theta_{\text{new}}, \mathbb{D}_{\text{new}}) =\; &(1-\lambda)\mathcal{L}_{\text{new}}(\theta_{\text{new}}, \mathbb{D}_{\text{new}}) +\nonumber\\ &\lambda\mathcal{L}_{\text{reg}}(\theta_{\text{base}}, \mathbb{D}_{\text{new}})\text{, where}\\
    \mathcal{L}_{\text{new}}(\theta_{\text{new}}, \mathbb{D}_{\text{new}}) = &\sum\limits_{b=1}^{B}\text{CE}(\mathbf{Y}_{b}, \hat{\mathbf{Y}}^{\text{new}}_{b})\text{, }\\
    \mathcal{L}_{\text{reg}}(\theta_{\text{base}}, \theta_{\text{new}}, \mathbb{D}_{\text{new}}) = &\sum\limits_{b=1}^{B}\text{KL}(\hat{\mathbf{Y}}^{\text{base}}_{b}, \hat{\mathbf{Y}}^{\text{new}}_{b})\text{, and}
\end{align}
\noindent
$\lambda$ is a factor that weights the contribution of $\mathcal{L}_{\text{new}}$ and $\mathcal{L}_{\text{reg}}$ to $\mathcal{L}_{\text{tot}}$, and $\text{KL}(a, b)$ is the KL-divergence between $a$ and $b$. We use $\lambda$ in order to balance the learning of the new information and the non-forgetting of the old information. The non-forgetting is implemented with the $\mathcal{L}_{\text{reg}}$, where the predictions of $\text{M}_{\text{new}}$ are sought to be as similar to the predictions of $\text{M}_{\text{base}}$. 

Finally, after calculating the $\mathcal{L}_{\text{tot}}$ for each sampling of data from $\mathcal{S}$, we obtain new optimized parameters for $\text{M}_{\text{new}}$ as

\begin{equation}
    \theta^{\star}_{\text{new}} = \underset{\theta_{\text{new}}}{\arg\!\min}\;\mathcal{L}_{\text{tot}}(\theta_{\text{base}}, \theta_{\text{new}}, \mathbb{D}_{\text{new}})\text{, }
\end{equation}
\noindent
where $\theta^{\star}_{\text{new}}$ are the new, optimized parameters. After obtaining $\theta^{\star}_{\text{new}}$, we update $\theta_{\text{new}}$ as
\begin{equation}\label{eq:final-update}
    \theta_{\text{new}} = \theta^{\star}_{\text{new}}\text{.}
\end{equation}
\noindent
Thus, $\text{M}_{\text{new}}$ is updated with the new information and also remembers old learned information, after applying~\eqref{eq:final-update}. The iterative process of our continual method for AAC is the process described by Equations~\eqref{eq:data-sampling} to~\eqref{eq:final-update}. The result of our method is the $\text{M}_{\text{new}}$ after the application of Eq.~\eqref{eq:final-update}.  

\section{Evaluation}\label{sec:evaluation}
In order to evaluate our method, we use a freely available and pre-optimized method as our $\text{M}_{\text{base}}$ and a freely available dataset different from $\mathbb{D}_{\text{ori}}$ to simulate $\mathcal{S}$, namely WaveTransformer (WT) and AudioCaps, respectively. $\mathbb{D}_{\text{ori}}$ used for WT is Clotho. We use mini-batches of size $B$ from AudioCaps to simulate $\mathbb{D}_{\text{new}}$, using only one epoch over AudioCaps. The performance of the continual learning is evaluated using metrics adopted usually in AAC task. Our code used for the implementation of our method can be found online\footnote{\url{https://github.com/JanBerg1/AAC-LwF}}. 

\subsection{Datasets and pre-processing}
Clotho~\cite{drossos2019clotho} is a freely available dataset for AAC, containing 3840 audio clips for training, 1046 for validation, and 1046 for evaluation. Each audio clip is of 15-30 seconds long and is annotated with five captions of eight to 20 words. This results to 19 200, 5230, and 5230 input-output examples for training, validating, and evaluating an AAC method, respectively. AudioCaps~\cite{audiocaps} is also a freely available AAC dataset, based on AudioSet~\cite{45857}. AudioCaps has 38 118 audio clips for training, 500 for validation, and 979 for testing. All audio clips are 10 seconds long, and clips for training are annotated with one caption while clips for validation and testing with five captions. These result to 38 118, 2500, and 4895 input-output examples for training, validating, and evaluating, respectively. In all experiments, as $\mathbb{D}_{\text{ori}}$ we use the training split of the corresponding dataset and as $\mathbb{D}_{\text{new}}$ the training split from the other AAC dataset. During the stage of hyper-parameter tuning we used as the validation split from $\mathbb{D}_{\text{ori}}$ and $\mathbb{D}_{\text{new}}$ to evaluate the performance of our method, while during testing we used the evaluation split as $\mathbb{D}_{\text{new}}$, from the corresponding dataset. These result to $K=19200$ for Clotho and $K=38118$ for AudioCaps. 

From all audio clips we extract $F=64$ log mel-band energies, using a 46 second long Hamming window with 50\% overlap. This results to $1292\leq T_{\text{a}}\leq2584$ for Clotho and $T_{\text{a}} = 862$ for AudioCaps. Additionally, for Clotho there are $8\leq T_{w}\leq20$ words in a caption and there are $W=4367$ unique words, while for AudioCaps there are $2\leq T_{w}\leq51$ words in a caption and there are $W=4506$ unique words. But, when $\text{M}_{\text{base}}$ is optimized on either Clotho or AudioCaps the $\text{M}_{\text{new}}$ is evaluated at the other dataset (i.e. $\text{M}_{\text{base}}$ trained on Clotho and $\text{M}_{\text{new}}$ evaluated on AudioCaps, and vice-versa). Since in our method we do not consider the case of learning new words, we keep only the common words from the dataset used for evaluation. For example, in the case of training on Clotho and evaluating on AudioCaps, we keep from AudioCaps only the words that exist in Clotho. The amount of words that we remove from AudioCaps is 1715. 

\subsection{M\textsubscript{base} model}
As M\textsubscript{base} we use the WT AAC model, presented in~\cite{tran2020wavetransformer}. WT consists of four learnable processes, three used for audio encoding and one for decoding the learned audio information to captions. WT takes as an input a sequence of audio features, e.g. $\mathbf{X}'$ or $\mathbf{X}$, and generates a sequence of words, e.g. $\mathbf{Y}'$ or $\mathbf{Y}$. Input audio features are processed in parallel by two different learnable processes, one for learning temporal patterns, $E_{\text{temp}}(\cdot)$, and one for learning time-frequency patterns, $E_{\text{tf}}(\cdot)$. $E_{\text{temp}}$ consists of 1D convolutional neural networks (CNNs), set-up after the WaveNet model~\cite{oord2016wavenet} and using gated and dilated convolutions. $E_{\text{tf}}$ is based on 2D depth-wise separable CNNs, capable to learn time-frequency information and proven to give state-of-the-art results in sound event detection~\cite{drossos:ijcnn:2020}. Both $E_{\text{temp}}$ and $E_{\text{tf}}$ do not alter the temporal resolution of their input and their output is concatenated and given as an input to a third learnable process, $E_{\text{merge}}(\cdot)$. $E_{\text{merge}}$ learns to intelligently merge the information from $E_{\text{temp}}$ and $E_{\text{tf}}$, producing as an output an encoded sequence of the input audio, containing both temporal and time-frequency information. 

The output of $E_{\text{merge}}$ is given as an input to a decoder, $D(\cdot)$ that is based on the Transformer model~\cite{vaswani2017attention}, using three stacked multi-head attention blocks. Each attention block takes as an input a sequence of tokens/words and uses two different multi-head attention processes. The first is a masked self-attention, for each token/word attending only to its previous ones in the input sequence. The second multi-head attention is a cross-modal attention, attending to the output of $E_{\text{merge}}$ given the output of the first, self-attention process. The first multi-head attention block $D$ takes as an input its outputs shifted right and applies a positional encoding. The output of the last multi-head attention block is given as an input to a classifier, which shares its weights through time and predicts the most probable word in each time-step of the output caption. WT is illustrated in Figure~\ref{fig:wt}, after~\cite{tran2020wavetransformer}. 

\begin{figure}
    \centering
    \includegraphics[width=\columnwidth]{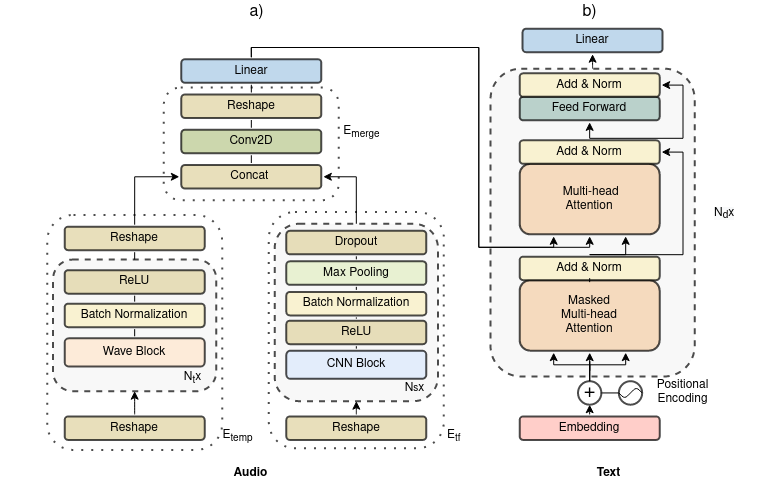}
    \caption{WT architecture, where a) is the encoder and b) the decoder, after~\cite{tran2020wavetransformer}.}
    \label{fig:wt}
\end{figure}

\subsection{Training, hyper-parameters, and evaluation}
We compare the performance of our proposed method against the following baseline scenarios: i) WT pre-trained on Clotho and evaluated on Clotho and AudioCaps, ii) WT pre-trained on AudioCaps and evaluated on Clotho and AudioCaps, and iii) WT pre-trained on Clotho, fine-tuned on AudioCaps, and evaluated on Clotho and AudioCaps. We term the above cases as $\text{WT}_{\text{cl-au}}$, $\text{WT}_{\text{au-cl}}$, and $\text{WT}_{\text{cl-ft}}$, respectively. For pre-training $\text{M}_{\text{base}}$, we use the training split of the corresponding dataset, employing the early stopping policy by using the corresponding validation split and the associated SPIDEr score. For both datasets we use 10 consecutive epochs for early stopping, detecting not improving SPIDEr score. As an optimizer we use Adam~\cite{kingma2014adam} with the proposed values for the hyper-parameters. Additionally, we use a temperature hyper-parameter at the softmax non-linearity of the classifier of $\text{M}_{\text{new}}$, as this has been found to improve the performance~\cite{li2017learning}. We use the value of 2 for this hyper-parameter. 

Using the above protocol, we evaluate the performance of our method using $\lambda = 0.70, 0.75, \ldots, 0.95, 1.0$ and $B=4, 8, 12$. We use the pre-trained WT on Clotho, and we simulate $\mathcal{S}$ as mini-batches of size $B$ from AudioCaps, as described by Eq.~\ref{eq:data-sampling}. We assess the performance of the $\text{M}_{\text{new}}$ at the 50th, 75th, and 150th update, and after using only once all data from AudioCaps, using SPIDEr score~\cite{Liu_2017}. SPIDEr~\cite{Liu_2017} is the weighted average of CIDEr and SPICE metrics. CIDEr~\cite{vedantam2015cider} employs weighted cosine similarity of $n$-grams, based on the term-frequency inverse-document-frequency (TFIDF), effectively quantifying the difference of the predicted and ground truth captions on using the same words to convey information. On the other hand, SPICE~\cite{anderson2016spice} analyzes the described scene and quantifies the differences of the predicted and ground truth caption in describing the same objects, attributes, and their relationships.

\begin{table}[!t]
\centering
\caption{SPIDEr score of the baseline scenarios\\}
\label{tab:baselines}
\begin{tabular}{lcc}
 \textbf{Baseline scenario} & \textbf{SPIDEr }$\pmb{\mathbb{D}_{\text{\textbf{ori}}}}$ & \textbf{SPIDEr }$\pmb{\mathbb{D}_{\text{\textbf{new}}}}$\\
\hline
$\text{WT}_{\text{cl-au}}$ & 0.182 & 0.108\\
$\text{WT}_{\text{au-cl}}$ & 0.318 & 0.102\\
$\text{WT}_{\text{cl-ft}}$ & 0.065 & 0.247
\end{tabular}
\end{table}

\section{Results}\label{sec:results}
In Table~\ref{tab:baselines} are the results of $\text{M}_{\text{base}}$, regarding the three different baseline scenarios. In Table~\ref{tab:results} are the obtained results of our method, for various values of $B$ and $\lambda$, focusing on the SPIDEr score for $\mathbb{D}_{\text{ori}}$ and $\mathbb{D}_{\text{new}}$. As can be seen from Table~\ref{tab:baselines} and from the cases of $\text{WT}_{\text{cl-au}}$ and $\text{WT}_{\text{au-cl}}$, the AAC method performs better on the $\mathbb{D}_{\text{ori}}$ than $\mathbb{D}_{\text{new}}$. This clearly shows that the model cannot perform equally well on the two different datasets, just by pre-training on one of them. Focusing on the $\text{WT}_{\text{cl-ft}}$, can be seen that the AAC method can perform good on the second dataset, i.e. $\mathbb{D}_{\text{new}}$, but the performance of the method on $\mathbb{D}_{\text{ori}}$ degrades considerably. This strengthens the need for our method, which aims at alleviating the degradation of performance on the $\mathbb{D}_{\text{ori}}$.

As can be seen from Table~\ref{tab:results}, it seems that the value of $B$ has an observable impact on the performance on $\mathbb{D}_{\text{ori}}$. That is, lower values of $B$ seem to not benefit the performance on $\mathbb{D}_{\text{ori}}$ for any value of $\lambda$. Specifically, for values of $B=4$, the SPIDEr score on $\mathbb{D}_{\text{ori}}$ is lower than the SPIDEr score for $\mathbb{D}_{\text{ori}}$ and for $B>4$, for any value of $\lambda$. The same stands true for $B=8$ and $B>8$. The above observation for $B$ suggests that the batch size for sampling the stream of data $\mathcal{S}$ can also act as a regularizer for the not-forgetting of information from the $\mathbb{D}_{\text{ori}}$. Regarding the impact of $\lambda$, one can directly see the effect of the $1-\lambda$ and $\lambda$ factors in Eq.~\eqref{eq:total-loss}, having $1-\lambda$ for scaling the effect of $\mathcal{L}_{\text{new}}$ and $\lambda$ for scaling the effect of $\mathcal{L}_{\text{reg}}$. Specifically, for $\lambda=1$ the SPIDEr score for $\mathbb{D}_{\text{new}}$ is lower than the SPIDEr score for $\mathbb{D}_{\text{ori}}$. This trend is in accordance with the observations from Table~\ref{tab:baselines}, and is an expected trend since the loss from $\mathbb{D}_{\text{new}}$ is turned to 0 for $\lambda=1$. Given the observations for $B$ from the same Table~\ref{tab:results}, it is indicated that using just the loss $\mathcal{L}_{\text{reg}}(\theta_{\text{base}}, \theta_{\text{new}}, \mathbb{D}_{\text{new}})$ for updating $\theta_{\text{new}}$ can enhance, up to an extent, the performance of the $\text{M}_{\text{new}}$ on the new data from $\mathcal{S}$. Similarly, for values of $\lambda<1.00$ the performance of $\text{M}_{\text{new}}$ on $\mathbb{D}_{\text{new}}$ increases for all values of $B$. Additionally, the value of $\lambda$ and the SPIDEr score on $\mathbb{D}_{\text{new}}$ have a reverse analogous relationship. 

In terms of better performing combination of $\lambda$ and $B$, we see two trends. There is the combination of $B=4$ and $\lambda=0.85$, which yields the best performance on $\mathbb{D}_{\text{new}}$ of SPIDEr$=0.230$. Additionally, there is the combination of $B=12$ and $\lambda=0.80$, which seems to act as the best regularizer for the performance on $\mathbb{D}_{\text{ori}}$, with SPIDEr$=0.186$. These results are in accordance with the previous observations for $B$ and $\lambda$, indicating some kind of trade-off for the values of $B$ and $\lambda$. Finally, comparing Tables~\ref{tab:baselines} and~\ref{tab:results}, one can see the benefit of our method, giving a good balance between the top performance on $\mathbb{D}_{\text{new}}$ and not deteriorating the performance on $\mathbb{D}_{\text{ori}}$. 

\begin{table}[!ht]
\centering
\caption{Results of continual learning using Learning without Forgetting for AAC, for various $B$ and $\lambda$. With bold are indicated the best SPIDEr scores for each dataset.\\}
\label{tab:results}
\begin{tabular}{cccc}
\textbf{batch size $\mathbf{B}$} & {{$\boldsymbol{\lambda}$}} & \textbf{SPIDEr }$\pmb{\mathbb{D}_{\text{\textbf{ori}}}}$ & \textbf{SPIDEr }$\pmb{\mathbb{D}_{\text{\textbf{new}}}}$\\
\hline
\multirow{7}{*}{4} & 0.70 & 0.098 & 0.239  \\
 & 0.75 & 0.102 & 0.215\\
 & 0.80 & 0.093 & 0.214\\
 & 0.85 & 0.115 & \textbf{0.230}\\
 & 0.90 & 0.133 & 0.215\\
 & 0.95 & 0.155 & 0.192\\
 & 1.00 & 0.163 & 0.119\\
 \hline
 \hline
 \multirow{7}{*}{8} & 0.70 & 0.113 & 0.210\\
 & 0.75 & 0.119 & 0.223\\
 & 0.80 & 0.132 & 0.220\\
 & 0.85 & 0.133 & 0.190\\
 & 0.90 & 0.156 & 0.187\\
 & 0.95 & 0.178 & 0.157\\
 & 1.00 & 0.165 & 0.114\\
 \hline
 \hline
 \multirow{7}{*}{12} & 0.70 & 0.109 & 0.211\\
 & 0.75 & 0.160 & 0.197\\
 & 0.80 & \textbf{0.186} & 0.157\\
 & 0.85 & 0.171 & 0.179\\
 & 0.90 & 0.182 & 0.153\\
 & 0.95 & 0.185 & 0.145\\
 & 1.00 & 0.176 & 0.115                 
\end{tabular}
\end{table}

\section{Conclusions}
\label{sec:conlusions}
In the paper we presented a first study of continual learning for AAC. Our method is based on the learning without forgetting method, which focuses on continuously updating the knowledge of a pre-trained AAC method on new AAC data, without degrading the performance of the AAC method on the originally used dataset during pre-training. For that reason, we employed a freely available and pre-trained AAC method and two freely available AAC datasets. We use the adopted AAC method which is pre-trained on one of the employed AAC datasets, and we use the other AAC dataset as a continuous stream of AAC data. We update the knowledge of the employed AAC method given the stream of AAC data. We compare our method against three baselines, two for training on one of the AAC datasets and evaluating on the other, and a third of training on one of the AAC datasets and fine-tuning the trained method to the other. Our results show that our method manages to not let the performance of the AAC method to deteriorate on the original AAC dataset, while, in the same time, manages to distil information from the new data to the employed AAC method. 

For future research, utilizing AAC datasets set in more distinct domains and training those in consecutive way to the model would provide more data on how effective these methods can be when used for AAC. Recent years continuous learning has been a hot issue and more methods have been introduced just during last few years, many of which might effective when utilized for AAC as well. 

\section{ACKNOWLEDGMENT}
\label{sec:ack}
The authors wish to acknowledge CSC-IT Center for Science, Finland, for computational resources. K. Drossos has received funding from the European Union’s Horizon 2020 research and innovation programme under grant agreement No 957337, project MARVEL.
\bibliographystyle{IEEEtran}
\bibliography{refs}
\end{sloppy}
\end{document}